\documentclass[10pt,twocolumn,letterpaper]{article}

\usepackage[pagenumbers]{wacv} 

\usepackage{graphicx}
\usepackage{amsmath}
\usepackage{amssymb}
\usepackage{booktabs}
\usepackage{algorithm}
\usepackage{algpseudocode}

\usepackage[pagebackref,breaklinks,colorlinks]{hyperref}

\usepackage[capitalize]{cleveref}
\crefname{section}{Sec.}{Secs.}
\Crefname{section}{Section}{Sections}
\Crefname{table}{Table}{Tables}
\crefname{table}{Tab.}{Tabs.}

\def\wacvPaperID{335} 
\def\confName{WACV}
\def\confYear{2025}

\begin{document}

\title{Divergent Domains, Convergent Grading: Enhancing Generalization in Diabetic Retinopathy Grading}

\author{Sharon Chokuwa \qquad Muhammad Haris Khan\\
Mohamed Bin Zayed University of Artificial Intelligence\\
United Arab Emirates\\
{\tt\small \{sharon.chokuwa, muhammad.haris\}@mbzuai.ac.ae}
}
\maketitle

\begin{abstract}
Diabetic Retinopathy (DR) constitutes 5\% of global blindness cases. While numerous deep learning approaches have sought to enhance traditional DR grading methods, they often falter when confronted with new out-of-distribution data thereby impeding their widespread application. In this study, we introduce a novel deep learning method for achieving domain generalization (DG) in DR grading and make the following contributions. First, we propose a new way of generating image-to-image diagnostically relevant fundus augmentations conditioned on the grade of the original fundus image. These augmentations are tailored to emulate the types of shifts in DR datasets thus increase the model's robustness. Second, we address the limitations of the standard classification loss in DG for DR fundus datasets by proposing a new DG-specific loss – domain alignment loss;  which ensures that the feature vectors from all domains corresponding to the same class converge onto the same manifold for better domain generalization. Third, we tackle the coupled problem of data imbalance across DR domains and classes by proposing to employ Focal loss which seamlessly integrates with our new alignment loss. Fourth, due to inevitable observer variability in DR diagnosis that induces label noise, we propose leveraging self-supervised pretraining. This approach ensures that our DG model remains robust against early susceptibility to label noise, even when only a limited dataset of non-DR fundus images is available for pretraining. Our method demonstrates significant improvements over the strong Empirical Risk Minimization baseline and other recently proposed state-of-the-art DG methods for DR grading. Code is available at 
\href{https://github.com/sharonchokuwa/dg-adr}{dg-adr}.
\end{abstract}

\section{Introduction}
\label{sec:intro}
\begin{figure}[t]
\centering
    \includegraphics[width=\linewidth]{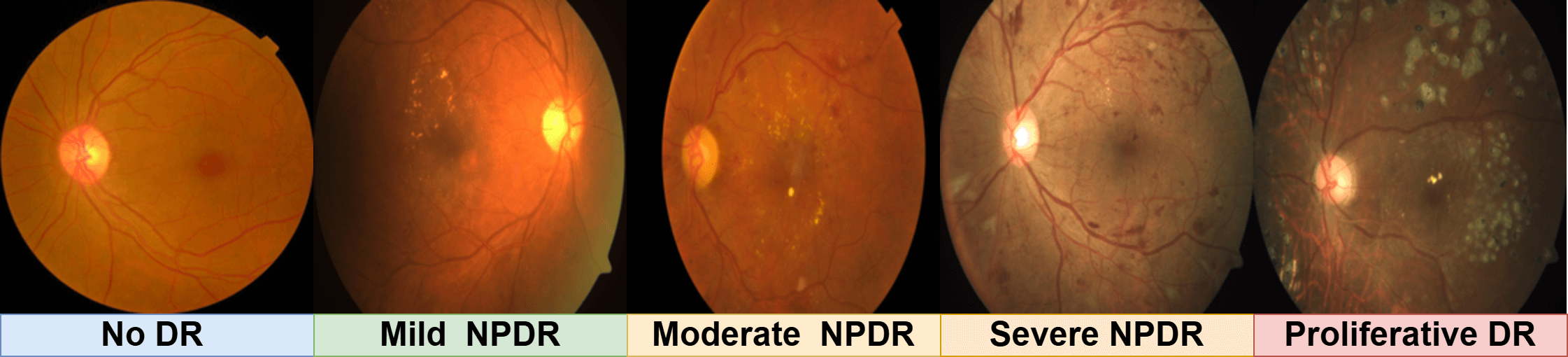}
    \caption{Illustration of DR Progression: From an initial healthy state to advanced stages of DR \cite{youtubeDeepLearning}. NPDR stands for non-proliferative Diabetic Retinopathy.}
    \label{fig:fundus_grades}
\end{figure}

Diabetic Retinopathy (DR) is a complication due to Diabetes Mellitus (DM). As of 2021, the global burden of DM surpassed 529 million people, a number projected to reach to a staggering 1.3 billion by 2050, with escalating prevalence rates worldwide~\cite{ong2023global}, thus inducing a parallel escalation in DR instances. DR manifests in several symptoms which include retinal hemorrhages, abnormal growth of blood vessels, aneurysm, hard exudates and cotton wool spots, with the severity grouped into five main progressive grades\slash classes (\cref{fig:fundus_grades}). Early DR detection is pivotal for enabling timely intervention since the initial DR stages commonly unfold without noticeable symptoms~\cite{kumar2016automated}.

The traditional approach to DR screenings, involving fundus scans analyzed by highly-skilled ophthalmologists, faces significant challenges due to manual process errors, subjective diagnosis, low ophthalmologist-to-patient ratios and prolonged examination turnaround times in some countries~\cite{technologyreviewGooglesMedical}, hindering efforts to address the rising incidences of DR within the DM population. In response to these challenges, various studies including \cite{dai2021deep,li2022deep,parthiban2023diabetic}, 
have focused on automating detection and grading using deep learning (DL) methods. However, these approaches face limitations, particularly in real-world deployment, as demonstrated by Google research team's DR solution, which performed well in controlled lab settings but failed upon deployment~\cite{technologyreviewGooglesMedical}.

This happens primarily due to the fact that DR fundus images suffer from shifts caused by the variation in acquisition procedures, diversity of population groups and dataset sampling. This domain shift, makes DR grading an even more challenging task since the grading already faces significant inter and intra-observer variability~\cite{atwany2022deep}. Our work aligns with the domain generalization (DG) research which aims to achieve robust model performance on DR out-of-distribution (OOD) fundus data \cite{atwany2022deep, galappaththige2024generalizing}. 

\begin{figure}[!h]
\centering
    \includegraphics[width=1.0\linewidth]{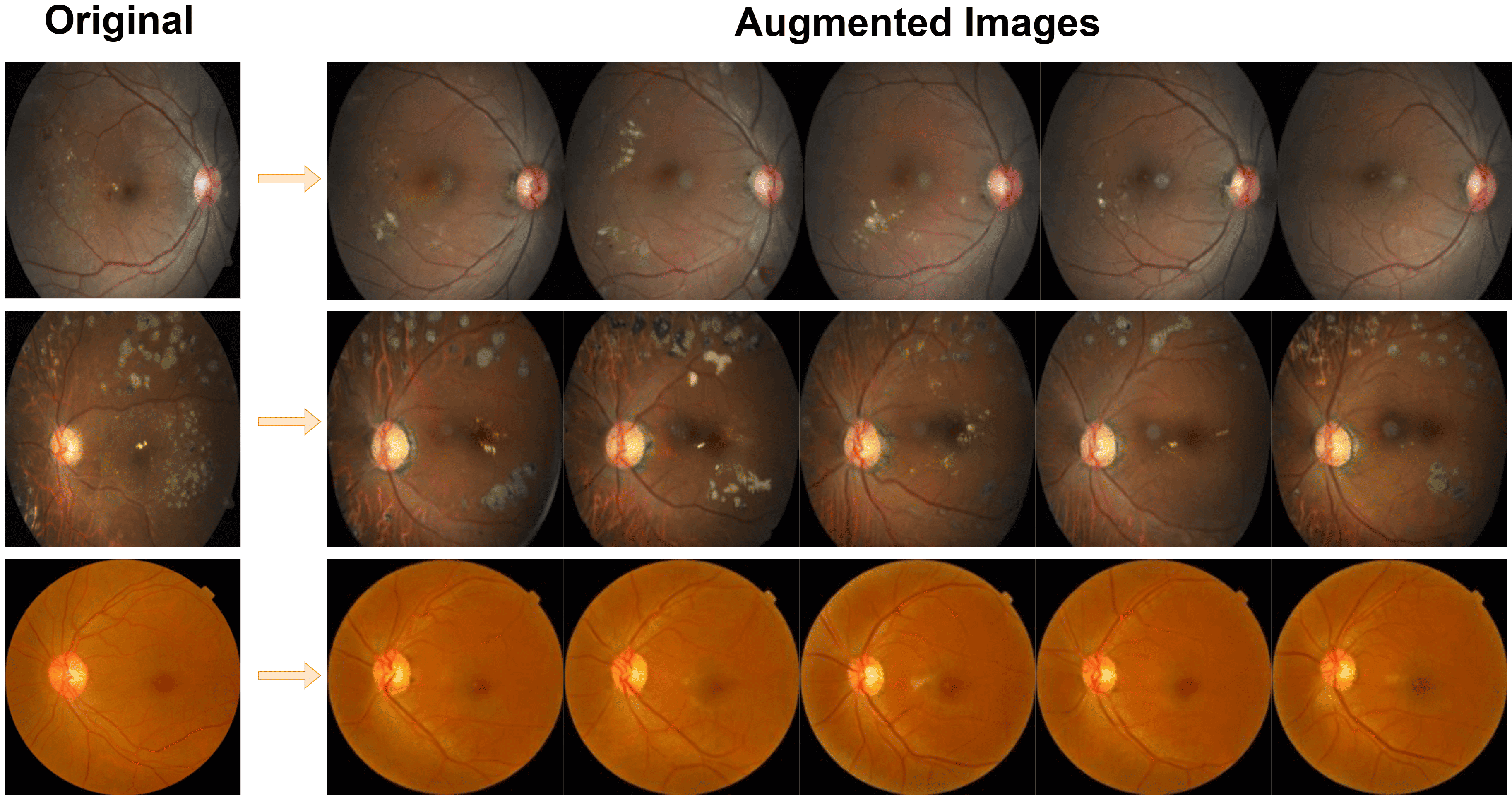}
    \caption{Samples of DR-Aug augmented fundus images, when the prompts used correspond to the original image's grade. First row corresponds to mild non-proliferative DR, the second row shows proliferative DR and third row no DR. The generated augmentations are consistent with the symptoms for the given text prompt and also exhibit some variations of the images present within the dataset (for that particular prompt), even when these variations are not within the original image itself. Temporal continuity for a particular DR grade is induced for a given original image \eg second row.}
    \label{fig:synthetic_images}
\end{figure}

While a few studies have been devoted to DG for DR over recent years, still considerable efforts are required for these models to attain satisfactory real-world performance. 
DR fundus datasets are plagued by limited temporal continuity, privacy concerns affecting data accessibility, high annotation costs and constrained ethnic diversity hindering generalizable performance on new data with different distributions. We therefore address these challenges by introducing a novel approach to augmenting images, which not only increases source domains' size at a relatively cheaper cost but also enhances semantic diversity, offering a solution which helps overcome domain shift in DR grading. \emph{We introduce domain generalizable DR fundus augmentations (DR-Aug)}, which produces varied and diagnostically relevant fundus images, that retain essential anatomical attributes while generating novel but clinically significant representations of diseased conditions. Notably, this capability is beyond what traditional augmentations (contrast adjustments, rotations etc.) can offer (see \cref{fig:synthetic_images}).

\begin{figure}[!t]
\centering
    \includegraphics[width=\linewidth]{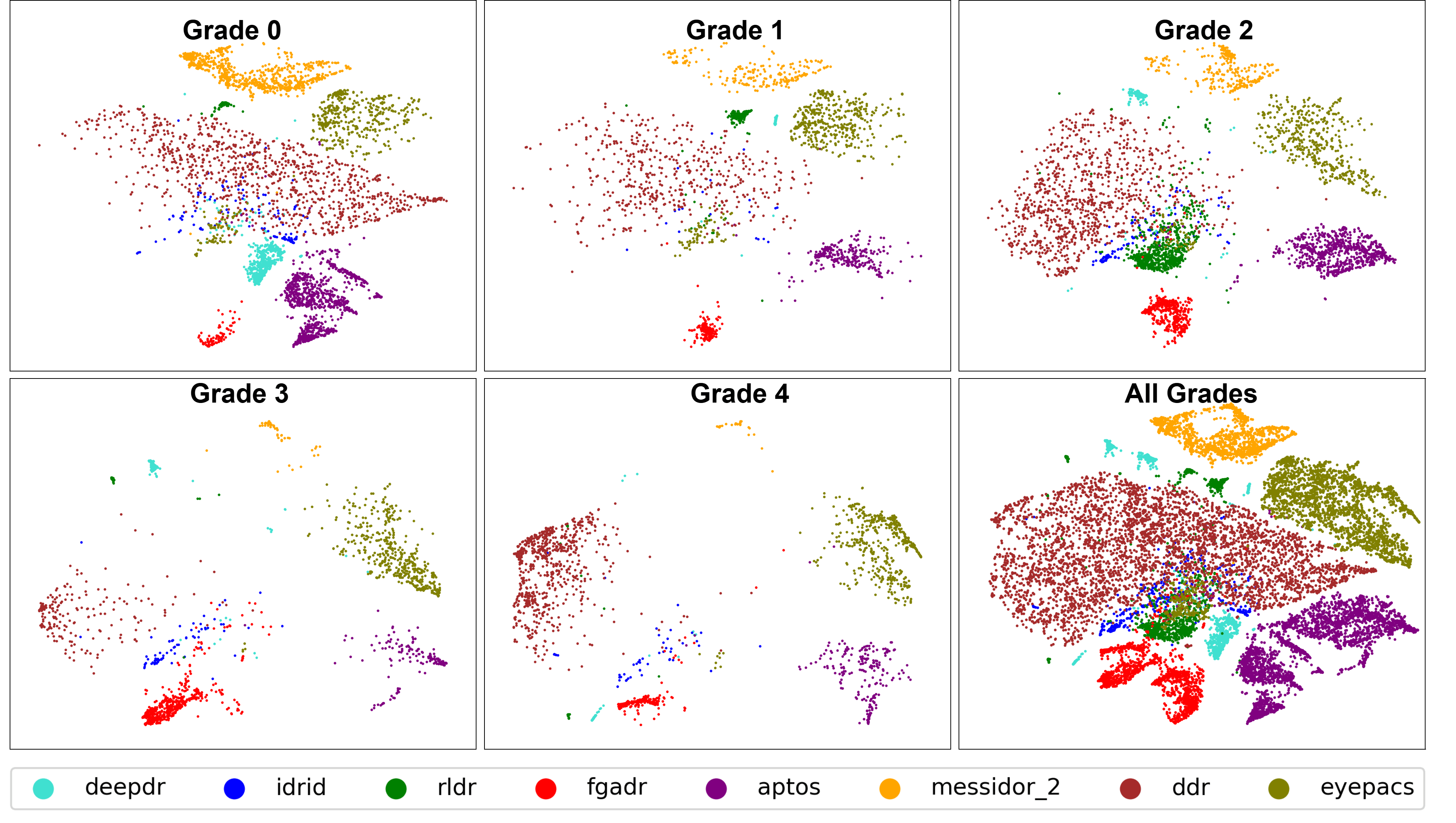}
    \caption{Representation of domain shift using t-SNE embeddings. Different colors correspond to distinct domains. Each plot represents a specific grade, ranging from 0 to 4. The final plot (right bottom) encapsulates the collective features of all grades within each respective domain.}
    \label{fig:domain_shift_tsne_classes}
\end{figure}

The feature space of the fundus datasets inherently reveals clustering based on domain-specific information and at the same time the intra-domain shift (variations within the same domain) is also evident---\cref{fig:domain_shift_tsne_classes}. As such ensuring a domain generalizable model requires scattering features across all domains, realizing a feature space capable of accommodating new OOD data. Accordingly, \emph{we propose a novel DG-specific loss, namely domain alignment loss (DomAlign).} This loss facilitates the aggregation of features corresponding to the same DR grade for all the source domains and simultaneously disperses the compactness of features originating from the same domain when they exhibit dissimilar semantic attributes, while accounting for potential intra-domain shift. Our DomAlign loss is flexible since it can be combined with any appropriate classification loss to demonstrate remarkable efficacy in ensuring consistent performance amidst domain shifts.

Furthermore, given the apparent high class imbalance in fundus datasets, it becomes crucial to devise a mechanism that effectively mitigates this imbalance. Consequently, \emph{we propose to leverage Focal loss \cite{lin2017focal} and integrate it with our DomAlign loss} to simultaneously tackle both the high class imbalance and the domain shift.
Finally, DR datasets demonstrate substantial observer variability, with inter-grader variability reaching up to 65\% and intra-grader variability up to 60\%~\cite{youtubeDeepLearning}, thereby introducing label noise. \emph{We therefore propose leveraging self-supervised learning (SSL) pretraining to reduce susceptibility to label noise} during the initial training phase, guiding the model to handle noisy labels more effectively. 

Our overall framework, featuring the non-trivial integration of aforementioned novel componenets, coined as  \emph{Domain-Generalized Augmentations for DR with Domain Alignment (DG-ADR)} consistently exceeds the efficacy of recently proposed DG methods by \emph{notable margins} across three different performance metrics. 

\section{Related Work}
\label{sec:lit_review}
\noindent\textbf{Domain Generalization for DR:}
Over recent years, a handful of studies including \cite{atwany2022drgen,che2023towards,chokuwa2023generalizing,wei2023caudr,matsun2023dgm,galappaththige2024generalizing}, have been put forth towards DG for DR. \cite{galappaththige2024generalizing}’s method enhances intermediate representations and mitigates overfitting by using a vision transformer backbone with self-distillation and adaptive prediction softening. However, most of the prior works were only evaluated on three or four datasets, constraining the genuine generalizability of such models in the face of more pronounced domain shifts. GDRNet \cite{che2023towards}, explored DG for DR across a broader spectrum of eight domains, a setting which we also follow. 

\noindent\textbf{Generative models in medical imaging:}
Some works have been proposed in the medical imaging domain, utilizing text-to-image generative models; these studies, including~\cite{chambon2022adapting,farooq2024derm,de2023medical}, were all evaluated in a single domain setting across chest X-rays, skin lesion images, MRI, and histopathology modalities. It is noteworthy that the aforementioned methods generate the medical images based on a text prompt only, in contrast, our focus in this study is on generating novel augmentations conditioned on both an existing fundus image and a text prompt, thereby minimizing the introduction of diagnostically irrelevant attributes. 

Our image-to-image augmentation strategy draws inspiration from the work by~\cite{trabucco2023effective}, which primarily focused on generative augmentation of natural images. In contrast, we aim to extend this paradigm to the more complex and demanding domain of DR datasets, since in the medical context, accurately generating subtle semantic attributes is of paramount importance compared to natural images. Furthermore, our work emphasizes evaluating these generated augmentations under domain shift conditions, which is a critical aspect not addressed by~\cite{trabucco2023effective}.

\noindent\textbf{Domain Alignment Loss:}
\cite{guo2023domain} proposed a domain-aware triplet loss specifically tailored for DG, aiming to prevent the model from preserving domain-specific information by dispersing features across domains while concurrently encouraging the clustering of features with similar image attributes. While our proposed alignment loss shares conceptual similarities with their work, a notable distinction lies in the underlying assumptions. \cite{guo2023domain}'s loss is formulated without considering intra-domain shift. In contrast, DR fundus datasets exhibit significant intra-domain shift, meaning the direct application of their loss inadvertently repels similar image attributes within sub-domains instead of attracting them. Our proposed DomAlign explicitly accounts for intra-domain shift, an aspect particularly pertinent in the context of medical datasets, ensuring more effective feature alignment within sub-domains.

\noindent\textbf{Self-supervised Learning:}
\cite{khanal2023improving} explored contrastive self-supervision and pretext task-based unsupervised approaches to mitigate the impact of noisy labels on histopathology and radiograph images. Their experiments validated the effectiveness of self-supervised learning pretraining in achieving robust generalization. Given that SSL pretraining is unaffected by noisy labels and thereby enhances model generalization in image classification tasks, we sought to explore this direction further. Different from \cite{khanal2023improving} and \cite{zheltonozhskii2022contrast}, we focus on a scenario where label noise is exacerbated by domain shifts. As the domains originate from diverse sources, the label noise becomes more pronounced and follows diverse distributions, making the classification problem significantly more challenging. While \cite{che2023towards} does not address this label noise, we consider tackling it to be crucial for improving model generalization.

\section{Method}

\noindent\textbf{Formal DG Setting:} Our approach follows the formal problem formulation outlined by~\cite{chokuwa2023generalizing,gulrajani2020search}, where they consider a set of $K$ source domains ($\{S_d\}_{d=1}^{K}$) during training. Each training domain $S_d = \{({x_i}^d, {y_i}^d)\}_{i=1}^{n}$ comprises samples drawn independently and identically from the distribution $p(X_d, Y_d)$, where $Y_d$ represents the label space and $X_d$ denotes fundus images. We construct mini-batches by sampling pairs $({x_i}^d, {y_i}^d)$ from all source domains. The goal in multi-source DG is to develop a model capable of accurate performance on a single, unseen target domain $T_d$ when the model is trained on $K > 1$ source domains.

\begin{algorithm}
\begin{algorithmic}[1]
  \State \textbf{Input:} $\langle \mathbf{x}', \mathbf{y}, \mathbf{d} \rangle$ from $\{S_d\}_{d=1}^{K}$ where $\mathbf{x}'$: original images, $\mathbf{y}$: class labels, $\mathbf{d}$: domain indices, synthetic image generator $h$, TrivialAugment $\phi$, network $f$, mini-batch size $M$, margin $\varepsilon$
  \State \textbf{Do:} Generate synthetic images $\mathbf{x}''$ = $h(\mathbf{x}', \mathbf{y})$
  \State \textbf{Init:} Initialize $f$ with SSL pretrained feature extractor $f_0$ 
  \For{mini-batch $(\mathbf{x}, \mathbf{y}, \mathbf{d})$}
    \State $\mathbf{x}''' \gets \phi(\mathbf{x})$, $\mathbf{x}\in\{\mathbf{x}', \mathbf{x}''\}$
    \State $\mathbf{\hat{y}} \gets f(\mathbf{x})$, $\mathbf{x}\in\{\mathbf{x}', \mathbf{x}'', \mathbf{x}'''\}$
    \State $\mathcal{L} = \frac{1}{M} \sum_{i=1}^{M} \ell_{\text{focal}}(\hat{y}_i, y_i) + \alpha\max\{0, \varepsilon + {D_p} - {D_n}\}$
    \State minimize $\mathcal{L}$
  \EndFor
  \State \textbf{Output:} Learned $f$ which yields $\mathbf{\hat{y}} = f(\mathbf{x}')$ for $T_d$
\end{algorithmic}
\caption{DG-ADR}
\label{algo:method}
\end{algorithm}

\subsection{Domain-generalizable Alignment Loss}
\label{subsec:proposed_method}
DR fundus datasets exhibit domain shifts, as demonstrated by the partitions of feature vectors within non-overlapping boundaries despite belonging to the same class, as illustrated in \cref{fig:domain_shift_tsne_classes}, resulting in the feature space being clustered based on domain information. We also assess the severity of the domain shift within DR domains by measuring the Kullback-Leibler (KL) divergences among the domain samples' features. The calculated KL divergences (given in Supplementary Material) also reveal notable shift among the domains. Our DomAlign loss focuses on reducing these inter-distances by scattering the features, ensuring that feature vectors from all domains belonging to the same grade converge onto the same manifold for better domain generalization at test time. 

Considering the presence of sub-domains within the primary DR domains, we aim to attract all samples with identical feature attributes. To achieve this, we define positive and negative samples as follows: positive samples, relative to a query sample in a mini-batch, are those with the same class label across all domains, including the domain of the query sample. In contrast, \cite{guo2023domain} defines positive samples without considering samples from the domain of the query sample, implicitly assuming the absence of intra-domain shift. Negative samples are those with the same domain but a different class label as the query sample. The concept of hard positive and hard negative samples is then introduced, referring to the $C$ nearest samples to the query sample with regard to either its class or domain labels.

DomAlign mines hard positive samples and hard negative samples at feature level within a mini-batch through attracting hard positives and repelling hard negatives within the same angular space where our other losses (e.g., the integrated Focal loss) also operates. For each feature vector $\mathbf{z}_i$ in a mini-batch, we initially form a weak positive set which comprises of all samples from the same class across all source domains $\{S_d\}_{d=1}^{K}$, including $\mathbf{z}_i$'s domain. Subsequently, we compute the cosine distance of this query sample $\mathbf{z}_i$ with all the other samples within its weak positive set. The top $C$ nearest samples are then selected as hard positives, denoted as set $P$. $C$ is experimentally chosen. The mean pooling of the hard positive distances yields the mean hard positive distance, denoted as $D_p$, as expressed in Equation \ref{eq:same_dist}. The mean hard negative distance, denoted as $D_n$ is obtained similarly as shown in Equation \ref{eq:diff_dist}, but over a hard negatives set $N$ - ($C$ nearest samples within the same domain as $\mathbf{z}_i$ but with different class labels). 

\begin{equation}
\label{eq:same_dist}
    {D}_p = \frac{1}{|{P}|} \sum_{j \in {P}} 1 - \frac{\mathbf{z}_i \cdot \mathbf{z}_j}{\|\mathbf{z}_i\| \cdot \|\mathbf{z}_j\|}
\end{equation}
\begin{equation}
\label{eq:diff_dist}
    {D}_n= \frac{1}{|{N}|} \sum_{k \in {N}} 1 - \frac{\mathbf{z}_i \cdot \mathbf{z}_k}{\|\mathbf{z}_i\| \cdot \|\mathbf{z}_k\|}
\end{equation}
Following Equation \ref{eq:same_dist} and \ref{eq:diff_dist}, our proposed DomAlign is computed as shown in Equation \ref{eq:align_loss}, with $\varepsilon$ as the margin of the loss hyperparameter:
\begin{equation}
\label{eq:align_loss}
    \mathcal{L_{\text{DomAlign}}} = \max\{0, \varepsilon + {D_p} - {D_n}\}
\end{equation}

Furthermore, Focal loss~\cite{lin2017focal} is introduced to counter the high class imbalance as it works by adjusting weights to prioritize the prone-to-misclassify instances. Our DG model is thus optimized on a combination of Focal loss and DomAlign, with $\alpha$ as a weighting parameter: $\mathcal{L} = \mathcal{L_{\text{Focal}}} + \alpha\mathcal{L_{\text{DomAlign}}}$. 

\begin{figure*}[!htp]
\centering
    \includegraphics[width=0.8\linewidth]{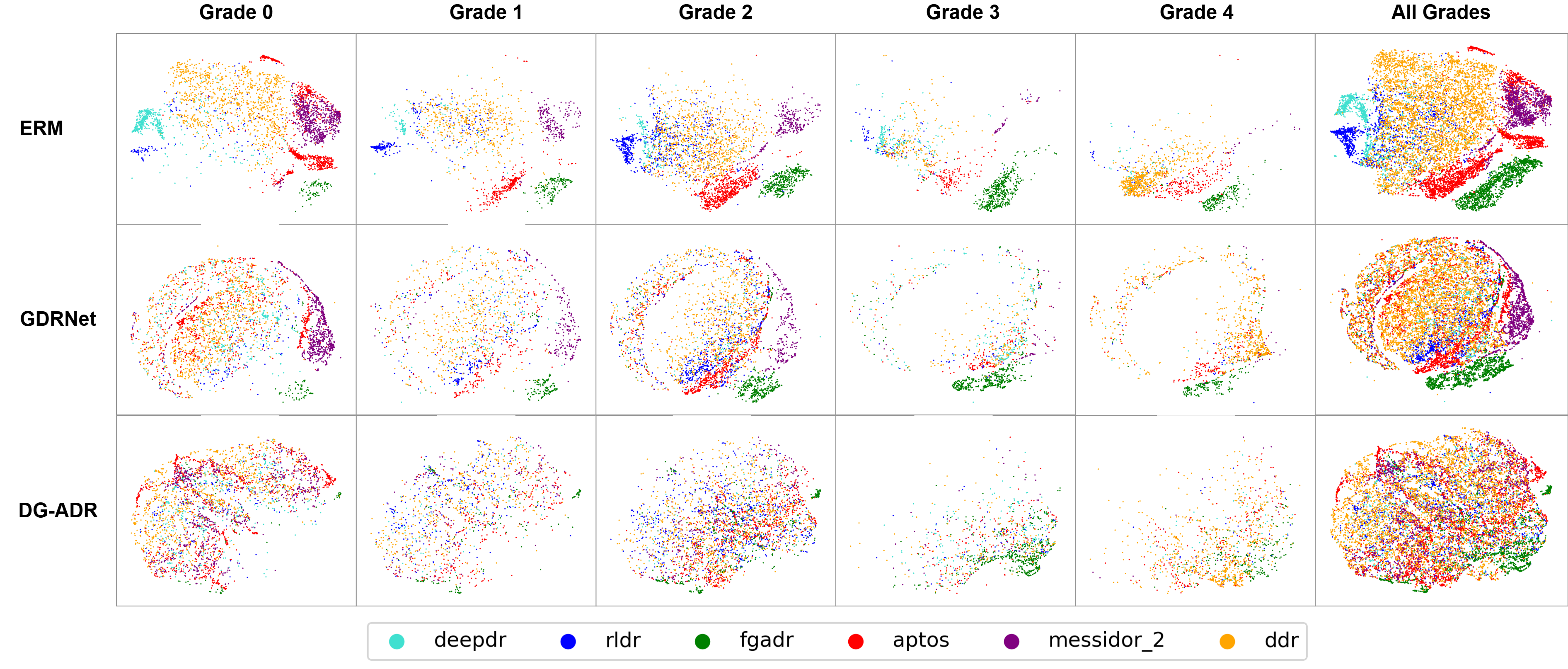}
    \caption{DG-ADR effectively scatters the feature space weakening the domain-specific information in comparison to ERM and GDRNet. The first two plots in each row depict t-SNE visualizations of two-dimensional features for each domain within a given grade. The final plot encapsulates the collective features of each domain regardless of class label (Grade 0 to 4).}
    \label{fig:compare_feature_space}
\end{figure*}

\subsection{DR Augmentations}
\noindent\textbf{DR-specific domain-level  variations:} Limited temporal continuity, privacy concerns affecting data accessibility, and high annotation costs contribute to the difficulty of training generalizable DR grading models. Our domain generalizable DR fundus augmentations (DR-Aug) tackles these challenges by generating novel DR images which maintain temporal continuity, increase the source domains data while also introducing synthetic yet relevant domain shift. 

\noindent\textbf{Our DR-Aug: } We utilize a pretrained Stable diffusion by \cite{rombach2022high}, which is a latent diffusion model, capable of generating an image from text description. The text (DR grade) is introduced as a control on the output images, ensuring alignment between the predicted noise and the desired output image upon noise subtraction. Inspired by \cite{meng2021sdedit} and \cite{trabucco2023effective}, we incorporate conditioning on both textual and image inputs such that during the reverse diffusion process, the original image is transformed into latent space and inserted at timestamp $t$, guiding the prediction of noise to align with the original image. 
Our attempts to directly generate fundus images from a text prompt only using a Stable Diffusion \cite{rombach2022high} model pretrained on natural images resulted in synthetic fundus images lacking meaningful information for DR detection. The work by \cite{trabucco2023effective}, finetunes the Stable diffusion model via Textual inversion only; our efforts to customize the model to our datasets the same way did not produce meaningful DR fundus images. Therefore, different from their approach we finetune the whole Stable diffusion model to ensure that the image generation aligns with our target data with higher granular semantic demands. We therefore, fine-tuned \cite{rombach2022high}'s image encoder using Dreambooth \cite{ruiz2023dreambooth}, which aims to customize the image generation by injecting the new subject (DR fundus images) onto the output target data via a unique concept by making both the text encoder and the U-Net trainable. We chose \textit{`a fundus image of $\langle$grade name$\rangle$'} as the instance prompt since the pretrained model does not have this concept in prior. 

Despite the shared objective of Dreambooth and Textual inversion \cite{gal2022image} methodologies, we manipulated both since their approaches are distinct. Our aim was to optimize the image encoder's finetuning process through Dreambooth and further refine the embeddings via Textual inversion. New word embeddings for each DR fundus grade were thus added to the model's vocabulary. The chosen text prompt was \textit{`a photo of fundus with $\langle$grade name$\rangle$'}, and via the new learned embeddings the model was able to distinguish between the different grades of DR. The subsequent phase involved generating augmentations based on the original image while ensuring that the augmented images remained within the same grade level as the original.
For each original image, N (which was randomly predetermined) augmentations are generated. Our DR-Aug model is represented as $h$ in~\cref{algo:method}. A sample of our augmented outputs are presented in \cref{fig:synthetic_images}, where we show that the resulting images closely align with the symptoms for the given text prompt.

\subsection{Conventional Augmentations} 
While DR-Aug increases our datasets and mainly produces images with semantic diversity that maintain temporal continuity and address shifts due to population diversity, domain shifts in DR datasets also stem from acquisition procedures and protocols. These include variations in lighting conditions, patient positioning, image compression, operator expertise, and other factors. To aid DR-Aug tackle these types of shifts, we incorporate TrivialAugment~\cite{muller2021trivialaugment} during training depicted as $\phi$ in \cref{algo:method}. TrivialAugment applies only a single uniformly sampled augmentation to each image with no tunable parameters. The range of augmentations employed include translation, blurring, equalize, Gaussian blurring, sharpness adjustments \textit{etc.}

\subsection{Self-supervised Pretraining}
The labeling of medical datasets, such as DR, is highly subjective, leading to label noise that significantly impacts the generalization capabilities of DL models \cite{youtubeDeepLearning}. Our objective is to address challenges stemming from this inevitable label noise. Unsupervised pretraining circumvents reliance on labels, enabling the feature extractor to learn robust features independent of noisy labels. By initializing our DG model with such pretraining, we aim to enhance its ability to learn robust features and reduce susceptibility to confusion caused by noisy labels.
 
In this study, we utilized publicly available DR fundus images originating from different sources, each of which was accompanied by only a single label. Consequently, we are unable to quantify the label noise; nevertheless, we acknowledge that label noise is inevitable and therefore exists in the provided DR fundus datasets. Therefore, an approach that can implicitly alleviate the impact of label noise will be favourable as it enables the development of a model with improved generalization capabilities. 
Given the shortage of alternative DR datasets, we explored other fundus datasets employed in evaluating various retinal diseases for the pretraining phase. We collected 16832 such images, resulting in a relatively small dataset since such datasets are scarce. The findings from \cite{newell2020useful}, indicate that the utilization of SSL pretraining, particularly when dealing with small datasets leads to improved performance. We selected the SimCLR framework proposed by \cite{chen2020simple} as our pretraining module; due to its simplicity and reliance on augmentations and contrastive loss which helps mitigate the adverse effects of label noise by focusing on learning generalizable features rather than overfitting to potentially noisy supervision. We initialized the SimCLR's image encoder $f_0$ using weights pretrained on ImageNet and subsequently utilized the weights from the SimCLR's image encoder $f_0$ to initialize our DG model $f$. See overview of our methodology in \cref{algo:method}.

\section{Experiments}

\subsection{Datasets}
\label{subsec:datasets}
\noindent\textbf{SSL Pretraining:}
For our SSL pretraining, we used various fundus datasets which are used for other fundus related illness including glaucoma detection, segmentation, cataract detection etc. These datasets are ORIGA \cite{zhang2010origa}, G1020 \cite{bajwa2020g1020}, ODIR-5K\cite{odir5k}, Drishti-GS \cite{sivaswamy2014drishti}, REFUGE \cite{tz6er97719}, RFMiD \cite{pachade2021retinal}, DIARETDB1 \cite{kauppi2007diaretdb1}, DRIONS-DB \cite{Carmona:2008:ION:1383660.1383874}, DRIVE \cite{drivedataset}, JSIEC \cite{cen2021automatic}, CHASE\_DB1 \cite{6224174}, Cataract dataset \cite{kaggleCataractDataset}, Glaucoma detection dataset \cite{kaggleGlaucomaDetection}, ROC \cite{niemeijer2009retinopathy} as well as DR1 and DR2 \cite{pires2014advancing}. The total for these datasets is 16832 images.
\noindent\textbf{Generative Model Finetuning:}
To avoid data leakage into the DG model, when finetuning the latent-diffusion model we only used the EyePACS \cite{kaggleDiabeticRetinopathy} dataset since it was the only dataset which had a fair number of images across all DR grades especially grade 4. Subsequently for DG training and evaluation, EyePACS was omitted.
\noindent\textbf{DG Model:}
DeepDR \cite{LIU2022100512}, Messidor-2 \cite{Messidor-2,10.1001/jamaophthalmol.2013.1743,decenciere2014feedback}, IDRID \cite{h25w98-18}, APTOS \cite{APTOS}, FGADR \cite{fgadr}, RLDR \cite{wei2021learn}, DDR \cite{LI2019} were employed with a leave-one strategy for training and testing our DG model. All these datasets include the 5 grades which are classified into grade 0 to 4. The sizes of the datasets are 1600, 1744, 516, 3656, 1842, 1593 and 12497, respectively. We present all dataset statistics and geographic origins in Supplementary Material.

\subsection{Experimental Setup}
\noindent\textbf{SSL pretraining:} We utilize a ResNet-50 pretrained on ImageNet as the image encoder, with our implementation building on the code by \cite{thalles_silva_2021_4486327}. \noindent\textbf{DR Augmentations:} For finetuning the diffusion image encoder, we utilize the Dreambooth implementation by \cite{von-platen-etal-2022-diffusers}. When employing Textual Inversion for fine-tuning the textual embeddings, our Dreambooth finetuned model serves as the initialization. We generate 250 synthetic images for each class per source domain (\textit{i.e.}, 1000 images per domain).

\noindent\textbf{DG training and evaluation:}
For a consistent implementation and fair evaluation we follow the DomainBed \cite{gulrajani2020search} protocol. We use the ResNet-50 architecture as the backbone for all our experiments, and keep the main hyperparameters consistent across methods, except for GDRNet, which adopts a batch size of 16 and training epochs of 100 due to observed performance degradation with larger batch sizes and longer training.  For consistency, we use the DR benchmark as proposed by \cite{che2023towards}. We provide detailed hyperparameter settings for all training phases in Supplementary Material and in our code.

\begin{table*}[t]
\scriptsize
\centering
\setlength{\tabcolsep}{3pt}
\resizebox{0.9\textwidth}{!}{
\begin{tabular}{ll|lllllllll}
\toprule
 &\textbf{Algo.} $\rightarrow$ &ERM \cite{gulrajani2020search} & GDRNet \cite{che2023towards} & DRGen \cite{atwany2022drgen} & Fishr \cite{rame2022fishr} & VAE-DG \cite{chokuwa2023generalizing} & SelfReg \cite{kim2021selfreg} & SD \cite{pezeshki2021gradient} & Mixup \cite{zhang2017mixup} & \textbf{DG-ADR} \\
\textbf{Domain} $\downarrow$ &  & &  &  &  &  &  &  &  &  \\
\midrule
 &Acc 
 & 39.9$\pm$0.4 
 & \textbf{49.2}$\pm$2.2 
 & 38.4$\pm$0.3
 & 40.4$\pm$3.7 
 & 33.2$\pm$2.3
 & 38.2$\pm$4.0 
 & 37.0$\pm$2.9 
 & 24.1$\pm$2.9 
 & 41.0$\pm$4.2 \\
DeepDR &AUC 
& 80.2$\pm$0.6 
& \textbf{84.3}$\pm$0.9 
& 79.5$\pm$0.9 
& 78.7$\pm$2.2 
& 73.2$\pm$1.2 
& 76.9$\pm$0.3 
& 79.5$\pm$1.1 
& 52.4$\pm$1.0 
& 80.6$\pm$1.2  \\
&F1 
& 35.6$\pm$1.1 
& \textbf{40.2}$\pm$1.4
& 34.9$\pm$0.7
& 37.4$\pm$2.1 
& 31.7$\pm$3.4 
& 29.6$\pm$3.0 
& 34.3$\pm$2.9 
& 11.6$\pm$1.7 
& 36.9$\pm$2.3 \\
\midrule
&Acc 
& 52.8$\pm$1.3 
& 52.5$\pm$2.1 
& 53.7$\pm$1.3 
& \textbf{55.5}$\pm$2.9 
& 53.4$\pm$1.1
& 54.3$\pm$2.2 
& 52.6$\pm$1.9 
& 40.7$\pm$0.8 
& 52.2$\pm$2.1 \\
IDRID  &AUC 
& 84.3$\pm$0.3 
& \textbf{86.2}$\pm$1.3 
& 85.3$\pm$0.7 
& 85.0$\pm$1.0 
& 81.7$\pm$2.4 
& 80.9$\pm$2.3 
& 85.4$\pm$0.3 
& 61.0$\pm$1.5 
& 85.1$\pm$0.7  \\
&F1 
& 42.1$\pm$3.5 
& \textbf{47.6}$\pm$2.1 
& 43.2$\pm$1.1 
& 45.7$\pm$1.0 
& 41.5$\pm$3.9 
& 40.8$\pm$1.1 
& 41.2$\pm$1.0 
& 19.7$\pm$0.7 
& 44.2$\pm$0.6 \\
\midrule
&Acc 
& 42.9$\pm$3.1 
& 41.7$\pm$1.8 
& 42.3$\pm$3.4
& 43.4$\pm$2.8 
& 39.7$\pm$4.5 
& 49.3$\pm$1.8 
& 43.9$\pm$4.9 
& 23.0$\pm$11.4 
& \textbf{51.9}$\pm$1.8 \\
RLDR  &AUC 
& 79.4$\pm$0.7 
& 81.0$\pm$0.9 
& 79.6$\pm$0.8 
& 79.5$\pm$1.3 
& 75.6$\pm$0.6 
& 77.8$\pm$0.2 
& 79.8$\pm$1.3 
& 53.0$\pm$1.3 
& \textbf{81.7}$\pm$0.3 \\
 &F1 
 & 32.3$\pm$2.5 
 & 32.1$\pm$1.2 
 & 33.8$\pm$3.4 
 & 34.9$\pm$2.3 
 & 32.5$\pm$3.6 
 & 35.3$\pm$0.8 
 & 34.6$\pm$3.2 
 & 13.7$\pm$6.1 
 & \textbf{40.0}$\pm$1.2 \\
\midrule
 &Acc 
 & 41.2$\pm$3.4 
 & 49.3$\pm$3.1 
 & 12.4$\pm$2.7 
 & 46.9$\pm$1.1 
 & 42.8$\pm$2.5 
 & 38.8$\pm$4.3 
 & 39.3$\pm$1.6 
 & 12.5$\pm$4.5 
 & \textbf{52.2}$\pm$1.8 \\
FGADR &AUC 
& 75.8$\pm$1.7 
& 79.6$\pm$1.0 
& 67.8$\pm$2.9 
& 76.6$\pm$0.5 
& 75.5$\pm$1.5 
& 74.6$\pm$0.5 
& 74.3$\pm$1.0
& 52.3$\pm$2.4 
& \textbf{80.8}$\pm$0.3 \\
&F1 
& 32.9$\pm$2.5 
& \textbf{42.1}$\pm$4.2 
& 7.5$\pm$1.6 
& 36.8$\pm$3.3 
& 34.5$\pm$1.5 
& 27.9$\pm$5.2 
& 30.6$\pm$1.8 
& 7.4$\pm$3.0 
& 41.2$\pm$3.1  \\
\midrule
 &Acc 
 & 59.5$\pm$3.0 
 & 58.5$\pm$3.7 
 & 65.1$\pm$2.6 
 & 60.4$\pm$3.1 
 & 44.9$\pm$1.0 
 & \textbf{70.6}$\pm$0.5 
 & 58.5$\pm$8.3
 & 51.8$\pm$1.8 
 & 69.7$\pm$2.0 \\
APTOS &AUC 
& 78.7$\pm$0.9 
& 78.7$\pm$1.1 
& 80.0$\pm$0.5 
& 79.8$\pm$0.3 
& 74.5$\pm$0.5
& 77.9$\pm$1.3 
& 78.9$\pm$1.4 
& 60.6$\pm$1.4 
& \textbf{80.4}$\pm$0.3 \\
&F1 
& 41.8$\pm$1.3 
& 40.6$\pm$1.9 
& 45.3$\pm$2.3 
& 44.2$\pm$1.1 
& 35.3$\pm$4.7 
& 44.4$\pm$1.6 
& 41.1$\pm$3.9 
& 22.6$\pm$1.9 
& \textbf{46.4}$\pm$0.1  \\
\midrule
 &Acc 
 & 61.5$\pm$0.7 
 & 61.5$\pm$0.6 
 & 61.4$\pm$0.3 
 & 61.1$\pm$0.7
 & 59.7$\pm$1.0
 & \textbf{63.4}$\pm$1.3
 & 61.5$\pm$1.1
 & 57.9$\pm$0.2 
 & 62.5$\pm$2.0  \\
Messidor-2 &AUC 
& 67.4$\pm$0.7 
& 69.4$\pm$1.9
& 66.3$\pm$0.7 
& 65.7$\pm$1.5 
& 63.9$\pm$2.3
& 66.4$\pm$0.7 
& 68.2$\pm$2.3 
& 48.9$\pm$2.0 
& \textbf{80.8}$\pm$0.8 \\
&F1 
& 33.2$\pm$3.7 
& 29.4$\pm$2.7 
& 30.9$\pm$1.4
& 29.0$\pm$1.6
& 25.4$\pm$5.8 
& 33.4$\pm$3.1
& 31.9$\pm$1.8 
& 15.0$\pm$0.2 
& \textbf{44.1}$\pm$0.8 \\
\midrule
&Acc 
& 58.3$\pm$0.3 
& 57.5$\pm$5.2 
& 59.3$\pm$2.7 
& 61.1$\pm$6.9 
& 50.8$\pm$2.9 
& \textbf{67.8}$\pm$0.8 
& 58.2$\pm$0.9 
& 42.7$\pm$2.0
& 65.0$\pm$0.2 \\
DDR &AUC 
& 80.6$\pm$0.4 
& 83.3$\pm$1.3
& 81.1$\pm$0.2
& 80.0$\pm$1.2 
& 79.0$\pm$0.7 
& 80.3$\pm$0.8
& 80.4$\pm$0.6 
& 50.3$\pm$0.7 
& \textbf{83.7}$\pm$0.8 \\
&F1 
& 44.0$\pm$1.0 
& 45.6$\pm$2.8 
& 44.4$\pm$1.0
& 43.7$\pm$2.4 
& 41.6$\pm$1.3
& 41.2$\pm$2.8 
& 44.5$\pm$0.9  
& 18.0$\pm$0.5 
& \textbf{49.5}$\pm$0.6 \\
\toprule
&Acc 
& 50.9$\pm$0.4 
& 52.9$\pm$1.5 
& 47.5$\pm$0.9  
& 52.7$\pm$1.4 
& 46.4$\pm$1.0 
& \underline{54.6}$\pm$1.5
& 50.2$\pm$1.7
& 36.1$\pm$2.1 
& \textbf{56.4}$\pm$0.2 \\
\textbf{Average}  &AUC 
& 78.1$\pm$0.1 
& \underline{80.4}$\pm$0.5 
& 77.1$\pm$0.5 
& 77.9$\pm$0.6 
& 74.8$\pm$0.7
& 76.4$\pm$0.6
& 78.1$\pm$0.8
& 54.1$\pm$0.5 
& \textbf{81.9}$\pm$0.4  \\
&F1 
& 37.4$\pm$0.9 
& \underline{39.7}$\pm$1.3 
& 34.3$\pm$0.4 
& 38.8$\pm$1.1 
& 34.7$\pm$0.8
& 36.1$\pm$1.7 
& 36.9$\pm$1.5 
& 15.4$\pm$1.2 
& \textbf{43.2}$\pm$0.6 \\
\bottomrule
\end{tabular}}
\caption{Main results --- comparison with other DG algorithms (Algo.) The given results for each target domain correspond to the average over three trials, with each model trained for 200 epochs. The standard deviations for each target domain are also given. \textit{Abbreviations}. Accuracy (Acc), AUC and F1-score (F1) in \%.}
\label{tab:main}
\end{table*}

\section{Results and Discussion}
\subsection{Quantitative Results} 
\noindent\textbf{Compared methods and evaluation metrics:} We set ERM \cite{gulrajani2020search} as our baseline, and also compare with other DG methods, which include DRGen \cite{atwany2022drgen}, Fishr \cite{rame2022fishr}, GDRNet \cite{che2023towards}, VAE-DG \cite{chokuwa2023generalizing}, SelfReg \cite{kim2021selfreg}, SD \cite{pezeshki2021gradient} and Mixup \cite{zhang2017mixup}. We employ three evaluation metrics: accuracy, area under the ROC curve (AUC) and F1-score and present our results as averages over three trials. A comparison between our proposed method DG-ADR and the ERM baseline, together with other DG algorithms is presented in \cref{tab:main}. 

\noindent\textbf{Results and discussion:}  Our DG-ADR outperforms all comparison algorithms across all three metrics, surpassing ERM by 5.5\%, 3.8\%, and 5.8\% in accuracy, AUC, and F1-score, respectively. Furthermore, compared to GDRNet, our approach achieves gains of 3.5\%, 1.5\%, and 3.5\% in accuracy, AUC, and F1-score, respectively. Our DG-ADR also performs best across the majority of the datasets (RLDR, FGADR, APTOS, Messidor-2 and DDR). The gains can be attributed to our DG-ADR being able to handle the DR shifts better by emulating the domain shift while consequently inducing temporal continuity via DR-Aug. Furthermore, the weakening of domain-specific bonds via DomAlign proves to render the embedding space robust to new OOD data. Using SSL pretrained weights does also significantly help to learn generalizable representations which are not misguided by noisy labels making the model more robust as evidenced by performance improvements, in comparison to the other methods which do not explicitly tackle this challenge. Our DG-ADR also shows an enhanced F1-score of 43.4\% which is significantly higher than all other methods, vindicating the better handling of the high class imbalance issues associated with DR fundus images via the employment of DomAlign with Focal loss. 

\begin{figure}[!htp]
\centering
    \includegraphics[width=\linewidth]{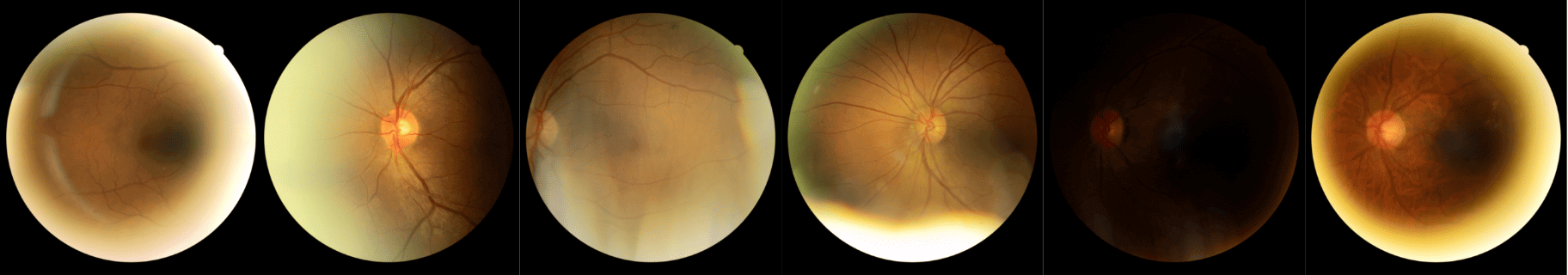}
    \caption{Samples of poor quality images in the DeepDR dataset. }
    \label{fig:deepdr_degraded}
\end{figure}

Across the domains, the DDR dataset showed the highest performance of 65.0\%, 83.7\% and 49.5\% in accuracy, AUC, and F1-score, respectively, when evaluated using our DG-ADR. This observation is due to the possibility that DDR is easier to classify given that its embeddings are highly dispersed across other domains' feature embeddings. Therefore, the application of DomAlign on DDR makes the learned features even more domain-agnostic. In contrast, DeepDR gave the least performance of 41.0\%, 80,6\% and 36.9\% in accuracy, AUC, and F1-score, respectively. This is due to highly degraded images present in this domain as illustrated in \cref{fig:deepdr_degraded}, which may make grading difficult even for trained experts. Finally, \cref{tab:dct_results} reports that our DG-ADR method, which explicitly addresses the intra-domain shift in DR fundus datasets, outperforms the approach by \cite{guo2023domain}, which does not account for this shift. Our method achieves nearly a 7\% improvement over their results.

\begin{table}[!ht]
\centering
\footnotesize
\setlength{\tabcolsep}{3pt}
\scalebox{0.7}{
\begin{tabular}{c|cccccccc}
\toprule
Method & APTOS & DDR & DeepDR & FGADR & IDRID & MESSIDOR-2 & RLDR & Avg. \\
\midrule
DCT \cite{guo2023domain} & 62.9$\pm$4.0 & 63.5$\pm$1.1 & 45.5$\pm$2.7 & 29.2$\pm$0.5 & 53.3$\pm$2.7 & 59.3$\pm$0.7 & 32.6$\pm$2.8 & 49.5$\pm$1.6 \\
\midrule
\textbf{Ours} & 69.7$\pm$2.0 & 65.0$\pm$0.2 & 41.0$\pm$4.2 & 52.2$\pm$1.8  & 52.2$\pm$2.1 & 62.5$\pm$2.0 & 51.9$\pm$1.8 & \textbf{56.4}$\pm$\textbf{0.2 }\\
\bottomrule
\end{tabular}
}
\caption{Comparison with DCT \cite{guo2023domain}.} 
\label{tab:dct_results} \vspace{-1em}
\end{table}

\subsection{Qualitative Results} 
\begin{figure}
    \centering
    \includegraphics[width=1.0\linewidth]{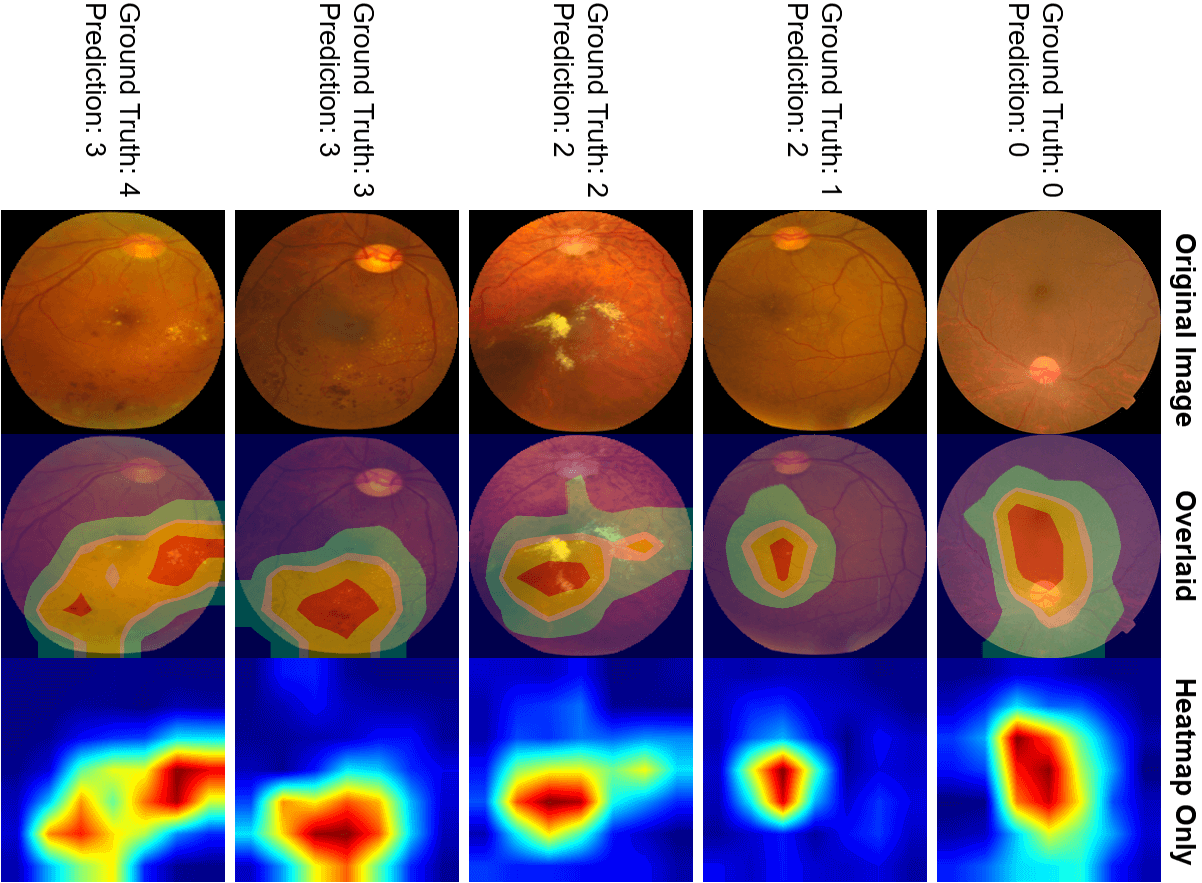}
    \caption{Grad-CAM heatmaps overlaid on original fundus DR images with APTOS as target domain. Ground truth and predictions generated by our DG-ADR model are given for comparison. GradCam identifies distinctive characteristic features pertaining to the image's grade. (Figure rotated 90 \textdegree clockwise)}
    \label{fig:heatmaps_gradcam} \vspace{-1em}
\end{figure}

We assess the efficacy of ERM, DG-ADR, and GDRNet in diminishing domain-specific information within the feature space. \cref{fig:compare_feature_space} illustrates the outcomes of feature extraction from each method using the t-SNE algorithm. Our findings demonstrate that our DG-ADR adeptly disperses features across the entire feature space, contrasting with ERM and GDRNet. Specifically, our method exhibits minimal clustering of domain information across all domains. Conversely, ERM displays significant clustering across nearly all domains, while GDRNet struggles to disperse Messidor-2, FGADR, APTOS, and DDR. These observations underscore the efficacy of DomAlign in effectively scattering features, creating a model which is robust to domain shift.
Our DR-Aug approach highlights that the augmented images (\cref{fig:synthetic_images}) maintain a close resemblance to the original images, while also presenting novel symptoms consistent with the original image's diagnosed DR grade. We utilized Grad-CAM \cite{selvaraju2017grad} heatmaps to visualize the regions of interest leading to classification predictions (Fig.~\ref{fig:heatmaps_gradcam}). Our findings indicate that the model focused on relevant symptoms corresponding to the classification predictions.

\begin{table*}[!ht]
\centering
\setlength{\tabcolsep}{3pt}
\label{tab:component_metrics}
\scalebox{0.63}{
\begin{tabular}{l|cccc|ccc|ccc|ccc|ccc|ccc|ccc|ccc}
\toprule
 &\multicolumn{4}{c|}{\textbf{Components}}& \multicolumn{3}{c|}{\textbf{DeepDR}} & \multicolumn{3}{c|}{\textbf{IDRID}} & \multicolumn{3}{c|}{\textbf{RLDR}} & \multicolumn{3}{c|}{\textbf{FGADR}} & \multicolumn{3}{c|}{\textbf{APTOS}} & \multicolumn{3}{c|}{\textbf{Messidor-2}} & \multicolumn{2}{c}{\textbf{Average}} \\ 
\midrule
& TriAug & DR-Aug & DomAlign & SSL& ACC & AUC & F1 & ACC & AUC & F1 & ACC & AUC & F1 & ACC & AUC & F1 & ACC & AUC & F1 & ACC & AUC & F1 & ACC & AUC & F1\\

\midrule
A. Focalloss & - & - & - &- & 29.5 & 78.0 & 27.1 & 48.3 & 81.9 & 42.0 & 33.7 & 76.8 & 29.3 & 44.8 & 73.6 & 33.0 & 54.0 & 77.5 & 37.8 & 60.7 & 69.9 & 32.6 &45.2 & 76.3 & 33.6 \\

B. A with TriAug & \checkmark & - & - &- & 41.3 & 80.9 & 33.6 & 45.2 & 85.0 & 41.8 & 33.1 & 79.3 & 31.8 & 47.8 & 74.8 & 33.8 & 41.4 & 78.2 & 35.7 & 58.4 & 69.3 & 28.9 & 44.5 & 77.9 & 34.3 \\

C. A with DR-Aug & - & \checkmark & - &- & 29.1 & 75.1 & 28.1 & 47.5 & 81.2 & 41.4 & 22.4 & 73.8 & 22.2 & 39.9 & 74.3 & 31.6 & 60.5 & 79.7 & 43.7 & 60.4 & 65.6 & 33.0 & 43.3 & 75.0 & 33.4 \\

D. A with B and C & \checkmark & \checkmark & - &- & 30.1 & 80.1 & 28.0 & 37.2 & 82.7 & 34.1 & 29.8 & 78.0 & 26.7 & 50.6 & 78.9 & 39.3 & 46.5 & 78.5 & 38.0 & 53.8 & 72.6 & 32.7 & 41.3 & 78.5 & 33.1 \\

E. D with DomAlign & \checkmark & \checkmark & \checkmark &- & 34.1 & 80.2 & 32.3 & 46.3 & 84.4 & 44.6 & 31.9 & 79.3 & 29.2 & 45.3 & 79.0 & 39.9 & 36.6 & 77.9 & 32.6 & 62.5 & 75.8 & 37.1 & 42.8 &79.4 & 35.9 \\

F. E with SSL & \checkmark & \checkmark & \checkmark & \checkmark & 31.1 & 81.4 & 30.9 & 47.5 & 84.9 & 41.1 & 33.9 & 80.1 & 31.4 & 48.7 & 77.5 & 41.3 & 69.2 & 80.9 & 46.0 & 64.3 & 74.1 & 42.4 & \textbf{49.1} &\textbf{79.8} & \textbf{38.9} \\
\bottomrule
\end{tabular}}%
\caption{Ablations studies over six target domains when the model is trained for 100 epochs. The reported results (in \%) are over a single trial. \textit{Abbreviations}. TriAug: TrivialAugment,  DR-Aug: DR augmentations, DomAlign: Domain Alignment, SSL: SSL pretraining.}
\label{tab:ablation_1} \vspace{-1em}
\end{table*}

\subsection{Ablation Studies} 
\noindent\textbf{Contribution of different components:} \cref{tab:ablation_1} presents the impact of individual components within our DG-ADR framework, including SSL pretraining, the incorporation of DR-Aug, and DomAlign, alongside the vanilla Focal loss model augmented with basic techniques. Ablation experiments were conducted across six domains with models trained for 100 epochs, owing to computational constraints. Nevertheless, a similar trend was observed when the model was evaluated across seven domains for 200 epochs. 

\begin{figure}[!h]
\vspace{-1em}
    \includegraphics[width=\linewidth]{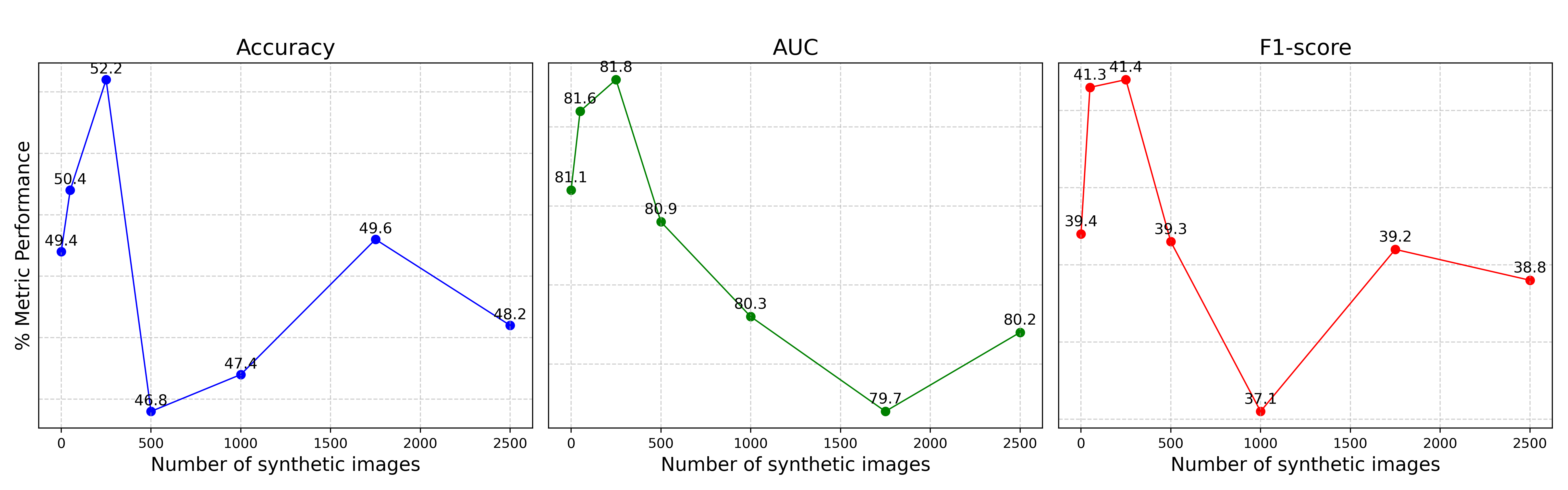}
    \caption{Effect of the image generation module (DR-Aug), which also illustrates of the effect of varying the number of synthetic images ($N$) added to each source domain. The presented results are the averages over six target domains (DeepDR, IDRID, RLDR, FGADR, APTOS and Messidor-2), and the model includes DomAlign and SSL pretraining. The highest performance is achieved when $N=250$ across all metrics.}
    \label{fig:effect_of_N}
\end{figure}

Our findings indicate that integrating DR-Aug along with TrivialAugment (Experiment D) enhances model performance by 3.5\% in AUC compared to the vanilla Focal loss (Experiment A) approach and the model employing only TrivialAugment (Experiment B). 
Experiment C which utilizes DR-Aug synthetic augmentations only in the absence of traditional augmentations exhibits a decreased performance in AUC. This is because while our DR-Aug synthetic augmentations are beneficial as shown in \cref{fig:effect_of_N}, their effectiveness is highly dependent on the ratio of synthetic to original images used. Specifically, in \cref{tab:ablation_1}, utilizing a high number of synthetic images to original images yields a lower performance in terms of AUC. This diminished performance can be attributed to the imbalance created by the excessive number of synthetic images. The results presented  \cref{tab:ablation_1} and our main experiments shown in \cref{tab:main} incorporate $N =1000$ (where $N=1000$ was initially randomly chosen) synthetic images per domain. However, experiments in \cref{fig:effect_of_N} demonstrate that using 1000 synthetic augmentations is sub-optimal, and the best results can be achieved using only 250 synthetic images. Given the relatively small size of the DR datasets used in this study, introducing 1000 synthetic images tends to overwhelm the original datasets by introducing synthetic biases.

Incorporating DomAlign (Experiment E) results in a more robust model against domain shift, evidenced by the improved performance metrics (with gains of 1.5\%, 0.9\% and 2.8\% in accuracy, AUC and F1-score, respectively). DomAlign's limited effectiveness on the APTOS dataset is likely due to stronger intra-domain shift in grade 0, yet still performing better than ERM and GDRNet in terms of clustering sparsity. 
Additionally, SSL pretraining (Experiment F) substantially enhances results across all metrics, with more gains in Accuracy and F1-score (\textit{i.e.}, gains of 6.6\% and 3.0\%, respectively), demonstrating that initializing the DG model with a checkpoint pretrained on fundus images leads to better generalization. Addressing noisy labels significantly assists the feature extractor in being less influenced by the presence of noise in the labels.

\noindent\textbf{With other losses:} In \cref{tab:ablation_2}, we study the effect of integrating DG-ADR with various classification losses. We show the results obtained when focal loss, cross entropy and weighted cross entropy (weights derivation discussed in Supplementary Material) are used. Results show that our framework significantly improves the model's performance despite the classification loss used.

\begin{table}[!htp]
\centering
\setlength{\tabcolsep}{4pt}
\label{tab:configurations_metrics}
\scalebox{0.8}{
\begin{tabular}{l|ccc}
\toprule
\textbf{Model Configuration} &ACC &AUC &F1 \\
\midrule
ERM  & 50.9 & 78.1 & 36.4 \\
ERM with DG-ADR & \textbf{54.7} & \textbf{79.5} & \textbf{40.5} ($\uparrow$) \\[0.2mm]
\hline
Weighted Cross Entropy   & 45.6 & 78.6 & 36.2 \\
Weighted Cross Entropy with DG-ADR  & \textbf{48.5} & \textbf{80.8} &\textbf{ 41.9} ($\uparrow$) \\[0.2mm]
\hline
Focal Loss  & 50.4 & 77.6 & 36.7 \\
Focal Loss with DG-ADR  & \textbf{52.6} & \textbf{81.2} & \textbf{42.0} ($\uparrow$) \\
\bottomrule
\end{tabular}}
\caption{Different loss configurations with DG-ADR, with the DG model initialized using ImageNet weights. The models were trained for 200 epochs and evaluated over seven domains. Results (in \%) are averaged over all the target domains on a single trial.}
\label{tab:ablation_2} \vspace{-1em}
\end{table}

\section{Conclusion}
We presented DG-ADR framework for cross-domain generalization in DR. It introduces DR-Aug, a (text+image)-to-image transformations model capable of generating novel diagnostically relevant fundus augmentations. 
It is followed by a new and effective DomAlign loss for combating the domain shift in DR fundus datasets.
Also, we tackle the high class imbalance by seamlessly integrating Focal loss with our DomAlign and also ensure better generalizability by leveraging SSL pretraining on non-DR fundus images to combat label noise induced by observer variability. Results yield significant gains by our method over other DG methods. 
Our DG-ADR is applicable to a broader range of tasks beyond DR grading, particularly when datasets encounter similar issues.

{\small
\bibliographystyle{ieee_fullname}
\bibliography{egbib}
}

\newpage
\twocolumn[{%
 \centering
 \large \textbf{Divergent Domains, Convergent Grading: Enhancing Generalization in Diabetic Retinopathy Grading}\\[1em]
 \large Supplementary Material \\[2.3em]
}]

\section{Quantifying the Domain Shift}
\begin{figure}[!h]
\centering
    \includegraphics[width=\linewidth]{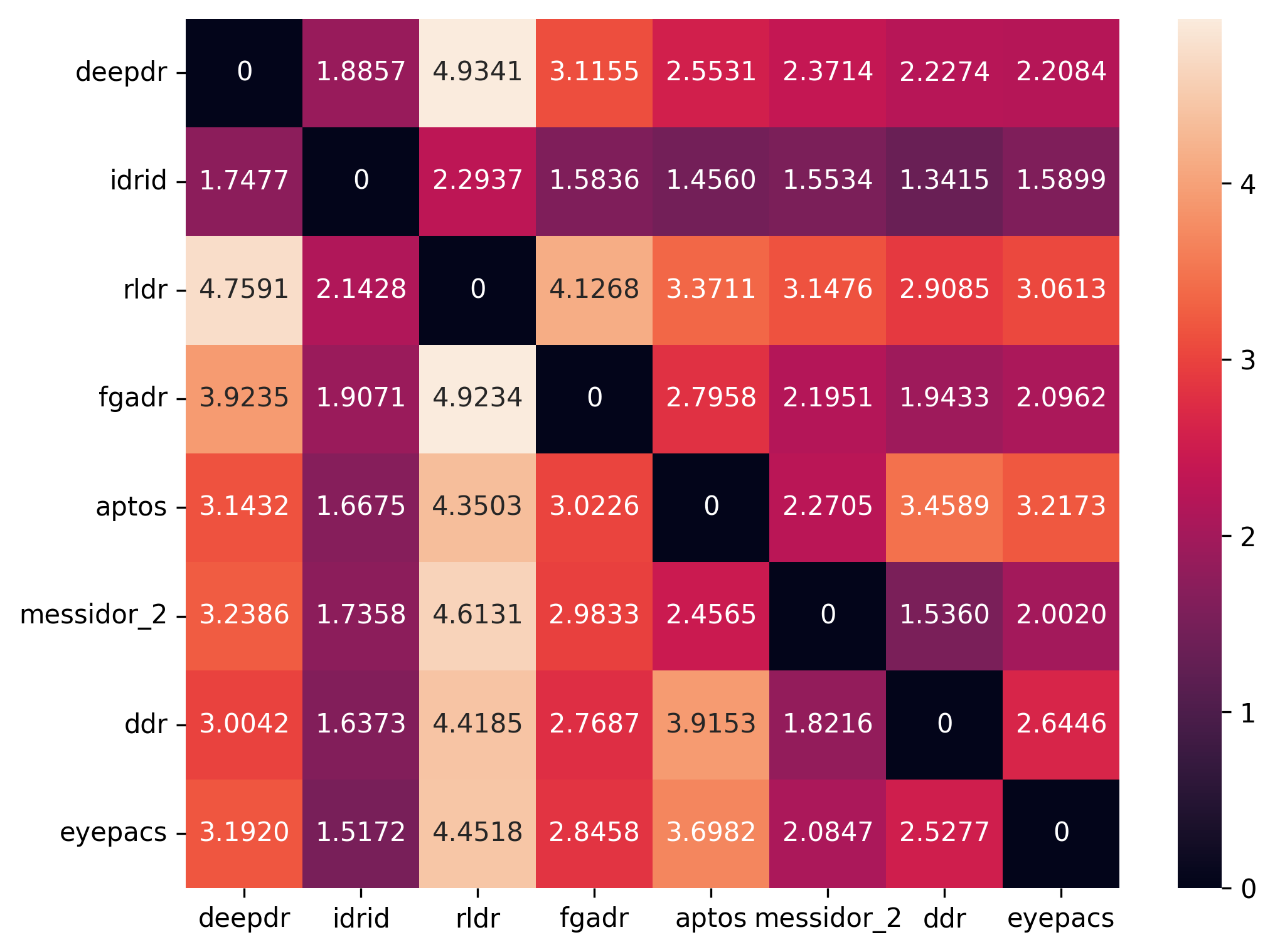}
    \caption{Heatmap visualization to show the KL divergences illustrating the extent of domain shift between domains. Lighter shades on the heatmap signify higher KL divergence, while darker shades indicate lower KL divergence.}
    \label{fig:kl_divergence_tsne_heatmap}
\end{figure}

To further quantify the domain shift empirically, we calculated the Kullback-Leibler (KL) divergences among the domains. Subsequently, a heatmap detailing pairwise KL divergences among all domains was constructed, as presented in \cref{fig:kl_divergence_tsne_heatmap}. Lighter shades on the heatmap indicate higher KL divergence, while darker shades signify lower KL divergence. The smallest domain shift is observed between the DDR \cite{LI2019} and IDRID \cite{h25w98-18} datasets, characterized by a KL divergence of 1.34, whereas the largest disparity is recorded between RLDR \cite{wei2021learn} and DeepDR \cite{LIU2022100512}, with a divergence of 4.93. This holistic analysis provides insights into the extent of domain shift observed between domains.

\section{Weighted Cross Entropy for DG}
\label{sec:weighted_ce}
The weights for our weighted cross entropy are calculated as follows: For each domain $S_d$, the inverse class weights are calculated as the ratio of total samples ($m_d$) in $S_d$ to the number of occurrences of each class $y$ in that domain $\omega_{yd} = \frac{m_d}{n_{yd}}$. The inverse domain weights are calculated as the ratio of the total number of samples $N$ to the number of samples in each domain $\theta_d = \frac{N}{N_d}$. The class weights are then normalized by dividing each weight by the maximum weight in that domain $\hat{\omega}_{yd} = \frac{\omega_{yd}}{\max(\omega_{yd})}$. The domain weights are normalized by dividing each weight by the maximum domain weight $\hat{\theta}_d = \frac{\theta_d}{\max(\theta_d)}$. Finally, the sample weights are calculated by multiplying the normalized class weights with the normalized domain weight for each class in each domain $w_{yd} = \hat{\omega}_{yd} \times \hat{\theta}_d$.

\section{Dataset Proportions}
The statistics of the datasets used for the SSL pretraining are presented in \cref{tab:ssl_datasets}. 
\begin{table}[!h]
\centering
\scalebox{0.9}{
\begin{tabular}{lr}
\toprule
Dataset                  & \multicolumn{1}{l}{Dataset Size} \\
\toprule
ORIGA                    & 650                              \\
G1020                    & 1020                             \\
ODIR-5K                  & 8000                             \\
Drishti-GS               & 101                              \\
REFUGE                   & 1200                             \\
RFMiD                    & 1200                             \\
DIARETDB1                & 89                               \\
DRIONS-DB                & 110                              \\
DRIVE                    & 40                               \\
JSIEC                    & 997                              \\
CHASE-DB1                & 28                               \\
ROC                      & 100                              \\
DR1 and DR2            & 2046                             \\
cataract\_dataset        & 601                              \\
Fundus\_Train\_Val\_Data & 650                              \\
\midrule
Total                    & 16832   \\
\bottomrule
\end{tabular}
}
\caption{Detailed breakdown of the compositions of the retinal datasets utilized during the SSL pretraining phase.}
\label{tab:ssl_datasets}
\end{table}

For our DG model the datasets utilised are presented in \cref{tab:ood_datasets}, which illustrates the dataset sizes as well as geographic origins.

\begin{table}[!h]
\centering
\scalebox{0.9}{
\begin{tabular}{lll}
\toprule
Dataset    & \multicolumn{1}{l}{Dataset size} & Dataset origin               \\
\midrule
DeepDR     & 1600                                   & Different hospitals in China \\
Messidor-2 & 1744                                   & France                       \\
IDRID      & 516                                    & India                        \\
APTOS      & 3656                                   & Rural India                  \\
FGADR      & 1842                                   & UAE                 \\
RLDR       & 1593                                   & USA                          \\
DDR        & 12497                                  & 23 provinces in China        \\
EyePACS    & 88698                                  & USA                          \\
\bottomrule
\end{tabular}
}
\caption{OOD datasets sizes and origins used for our DG model.}
\label{tab:ood_datasets}
\end{table}

\begin{figure*}[!htp]
\centering
    \includegraphics[width=0.8\linewidth]{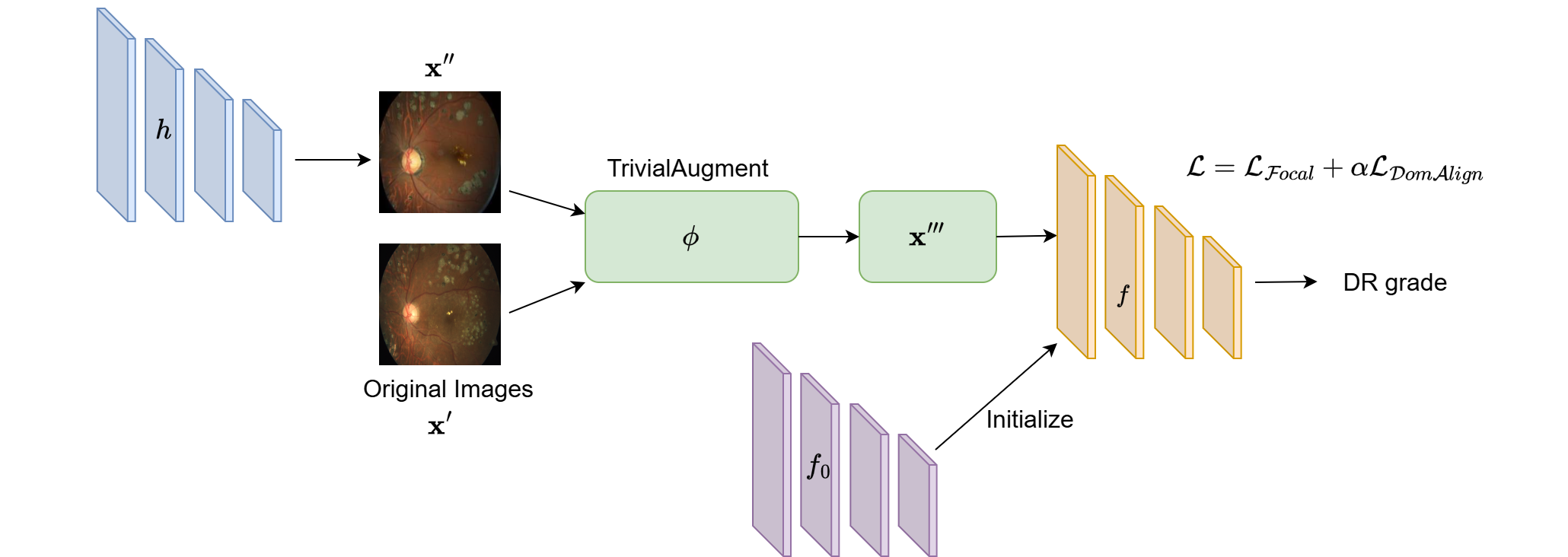}
    \caption{Overview of our method}
    \label{fig:method}
\end{figure*}

\section{Additional Implementation Details}
\noindent\textbf{SSL pretraining:} We employ a batch size of 128. The model is trained for 200 epochs, utilizing a learning rate of 0.0003 and a weight decay of 1e-4.

\noindent\textbf{DR Augmentations:} Our Dreambooth utilized the hyperparameters outlined in \cref{tab:dreambooth_hyperparametes}. When Textual Inversion finetuning we utilize a training batch size of 1, a learning rate of 0.0005, and a maximum of 1000 training steps. Consistently, we employ the same instance and class prompts as employed in Dreambooth fine-tuning, maintaining a real guidance strength of 0.5, a Stable Diffusion guidance scale of 7.5, conducting 1000 denoising steps and 200 inference steps. Our code builds on \cite{trabucco2023effective} for generating the images.

\noindent\textbf{DG training and evaluation:}
Tunable parameters $\varepsilon$, $C$ and $\alpha$ were experimentally chosen to be 0.1, 5 and 10, respectively. The batch size is configured as 128, with training conducted over 200 epochs (for our main results), employing a learning rate of 0.001 and SGD optimizer.

\begin{table}[!htbp]
\centering
\scalebox{0.9}{
\setlength{\tabcolsep}{1pt}
\begin{tabular}{@{}l|l@{}}
\toprule
\textbf{Hyperparameter} & \textbf{Value} \\ \midrule
Pretrained model  & CompVis/stable-diffusion-v1-4 \\
With prior preservation & True \\
Prior loss weight & 1.0 \\
Instance prompt & \textit{`a photo of fundus with $\langle$grade name$\rangle$'} \\
Class prompt & \textit{`a photo of fundus'}\\
Resolution & 512 \\
Train batch size & 2 \\ 
Learning rate & 5e-6 \\ 
Learning rate scheduler & constant \\ 
Max train steps & 30000 \\ 
Number class images & 500 \\ 
\bottomrule
\end{tabular}
}
\caption{Finetuning Dreambooth hyperparameters and their values.}
\label{tab:dreambooth_hyperparametes}
\end{table}

\section{Generative model finetuning data}
\textit{What implications arise from incorporating target data in finetuning the latent diffusion model?} To ensure that the datasets used in the OOD evaluations are distinct from the ones used in finetuning the latent diffusion model, only the EyePACS dataset was used for the latter purpose. This dataset was chosen due to its expansive size and rich variety of image representations across all grades, particularly grade 4 compared to the other domains, ensuring an ample size of samples for the fine-tuning process. We required approximately 200 images for training and 500 images as class images in each grade's Dreambooth model. Our rationale for this approach was to ensure that there is no data leakage between the diffusion model and the DG model, since in a DG setting, we assume that the target data is completely unseen throughout the training process.

\end{document}